\documentclass{aastex}

\begin{document}


\title{Quasars in the 2MASS Second Incremental Data Release}


\author{Wayne A. Barkhouse and Patrick B. Hall\altaffilmark{1}}
\affil{Department of Astronomy, University of Toronto,\\ 60 St. George Street, 
Toronto, ON, M5S 3H8}

\email{barkhous@astro.utoronto.ca}


\altaffiltext{1}{Current Affiliations:
Princeton University Observatory, Princeton, NJ 08544-1001 and
Pontificia Universidad Cat\'{o}lica de Chile, 
Departamento de Astronom\'{\i}a y Astrof\'{\i}sica, 
Facultad de F\'{\i}sica, Casilla 306, Santiago 22, Chile;
E-mail: phall@astro.puc.cl}


\begin{abstract}

Using the 2MASS Second Incremental Data Release, we have searched
for near infrared counterparts to 13214 quasars from the
\citet{vv00} catalog. We have detected counterparts
within $4\arcsec$ for 2277 of the approximately 6320 
quasars within the area covered by the 2MASS Second Incremental Data Release.
Only 1.6\% of these are expected to be chance coincidences. 
Though this sample is heterogeneous, we find that known radio-loud quasars are 
more likely to have large near-infrared-to-optical luminosity ratios than
radio-quiet quasars are, at a statistically significant level.
This is consistent with dust-reddened quasars being more 
common in radio-selected samples than in optically-selected samples,
due to stronger selection effects against dust-reddened quasars in the latter.
We also find a statistically significant dearth of optically luminous quasars 
with large near-infrared-to-optical luminosity ratios.  
This can be explained in a dust obscuration model but not in a model where
synchrotron emission extends from the radio into the near-infrared and creates
such large ratios.
We also find that selection of quasar candidates from the $B-J/J-K$ color-color
diagram, modelled on the $V-J/J-K$ selection method of 
Warren, Hewett \& Foltz (2000), is likely to be more sensitive to 
dust-obscured quasars than selection using only infrared-infrared colors.

\end{abstract}


\section{Introduction}\label{Intro}

Active Galactic Nuclei (AGN) have been studied for several decades,
but our understanding of many aspects of AGN is still rudimentary.
The observed emission from AGN is thought to be powered by the release of
gravitational potential energy from matter falling into supermassive
black holes, but the details of this process and their effect 
on the observed spectra of AGN are not understood. Our inability to
definitively understand AGN to date may be due in part to selection effects.
Most high-luminosity AGN (quasars) have been selected using their
rest-frame ultraviolet through optical colors in optically
magnitude-limited samples. It is therefore almost certain
that dust obscuration has masked some portion of the AGN population
from detection, and that AGN found by their ultraviolet-optical colors
are less dusty than the truly typical AGN \citep{web95,bak97,wh97}.
This bias against dust-obscured objects can affect our understanding of AGN,
in particular their connection to their host galaxies and nuclear environments.
(We use the term dust {\em obscuration} to refer to the combined effects of
dust {\em extinction}, the reduction in observed flux from an object screened 
by dust, and dust {\em reddening}, which arises from typical `non-gray' 
extinction that is stronger at bluer wavelengths.  Extinction can occur
without reddening, but reddening cannot occur without extinction.)

The Two-Micron All-Sky Survey \citep[2MASS,][]{skr97} has now made it feasible
to select samples of AGN in the near infrared (NIR). A sample selected
only by NIR magnitude is an excellent way to reduce 
any bias against dust-obscured objects, but for efficient
selection of AGN candidates it is usually necessary to use additional
selection criteria such as morphology and color. To aid in this endeavor
and to provide a list of AGN with homogeneous infrared data, in this paper
we present the results of a cross-correlation of the 2MASS Second Incremental
Data Release with the quasar catalog of \citet[hereafter VV00]{vv00}.

\section{Catalog Cross-Correlation}\label{Catalog}

We took all 13214 quasars (defined as AGN more luminous than $M_B=-23$,
regardless of radio power) from Table 1 of VV00 and searched for unresolved 
2MASS counterparts within $5\arcsec$.
We found 2327 counterparts for 2320 quasars and discarded 
the more distant ones for the seven quasars with two counterparts. 
There were 4000 additional quasars
which fell within the 2MASS areal coverage but were not detected by it.

The internal errors on the 2MASS astrometry are on average 0\farcs3
for our detections, so the uncertainty in the astrometric matching
is dominated by the errors on the quasar coordinates. Only 53 of the
VV00 quasars 
have positions which may be uncertain by several
arcminutes (denoted `A' by VV00).  The majority (8884) have optical
positions which are accurate to $\lesssim 1 \arcsec$ (`O'), and 523
have similarly accurate radio positions (`R'). The remaining 3754
have positions which should be good to a few arcsec (these have no code
in VV00, but we denote them here by `N'). 

A search for counterparts to fake objects indicates that
even our near-IR counterparts of code `N' objects are likely to be real.
We shifted the declinations of all 13214 VV00 quasars 6$\arcmin$ to the north
and searched for matches within $5\arcsec$ to determine the number of spurious
counterparts in our sample.  Only 74 fake-object counterparts were found, as
opposed to 2320 real-object counterparts. This indicates that there are
few spurious counterparts in our sample, but the contamination can be reduced
even further by requiring a match within $4\arcsec$ instead of $5\arcsec$.
Only 43 of 2320 counterparts (including the one code `A' object among the 2320)
have a separation between the VV00 and 2MASS positions of $>4\arcsec$, while
37 of 74 fake-object matches have separations $>4\arcsec$. Thus the number of
counterparts to VV00 quasars with separations between $4\arcsec$ and $5\arcsec$
is consistent with the number expected from random chance, and for our final
sample we only consider counterparts with separations $<4\arcsec$.
Our final catalog consists of 2MASS counterparts to 2277 
VV00 quasars, of which about 37 (1.6\%) are expected to be chance coincidences.
The 2MASS and VV00 identifications of these 2277 quasars are given in 
Table~\ref{tab_detected}. 
The 4000 VV00 quasars not detected by 2MASS despite lying within the area of 
the Second Incremental Data Release are listed in Table~\ref{tab_missing}.

\section{Properties of Quasars Detected by 2MASS} \label{resultdis}

Radio luminosities were computed for all VV00 quasars that had measured
6 cm radio flux densities.  We assumed a radio spectral index of 
$\alpha = -0.5$, a Hubble constant of 
$H_{o}=50~\mbox{km}~\mbox{s}^{-1}~\mbox{Mpc}^{-1}$ 
and $q_{o}= 0.0$. We classified each quasar as radio-loud (RLQ) or radio-quiet 
(RQQ) based on the ratio of its radio to optical luminosity ($L_{rad}/L_{opt}$).
The divide between RLQ and RQQ was taken as log($L_{rad}/L_{opt}$)=0.
Optical luminosities were calculated from $M_{B}$ as given in VV00.
Near infrared luminosities 
were calculated from $K$ magnitudes given in the 2MASS catalog 
using $k$-corrections for a power-law continuum with $\alpha=-0.5$,
the same value used by VV00 to calculate their $M_B$ values.

\begin{figure}
\plotone{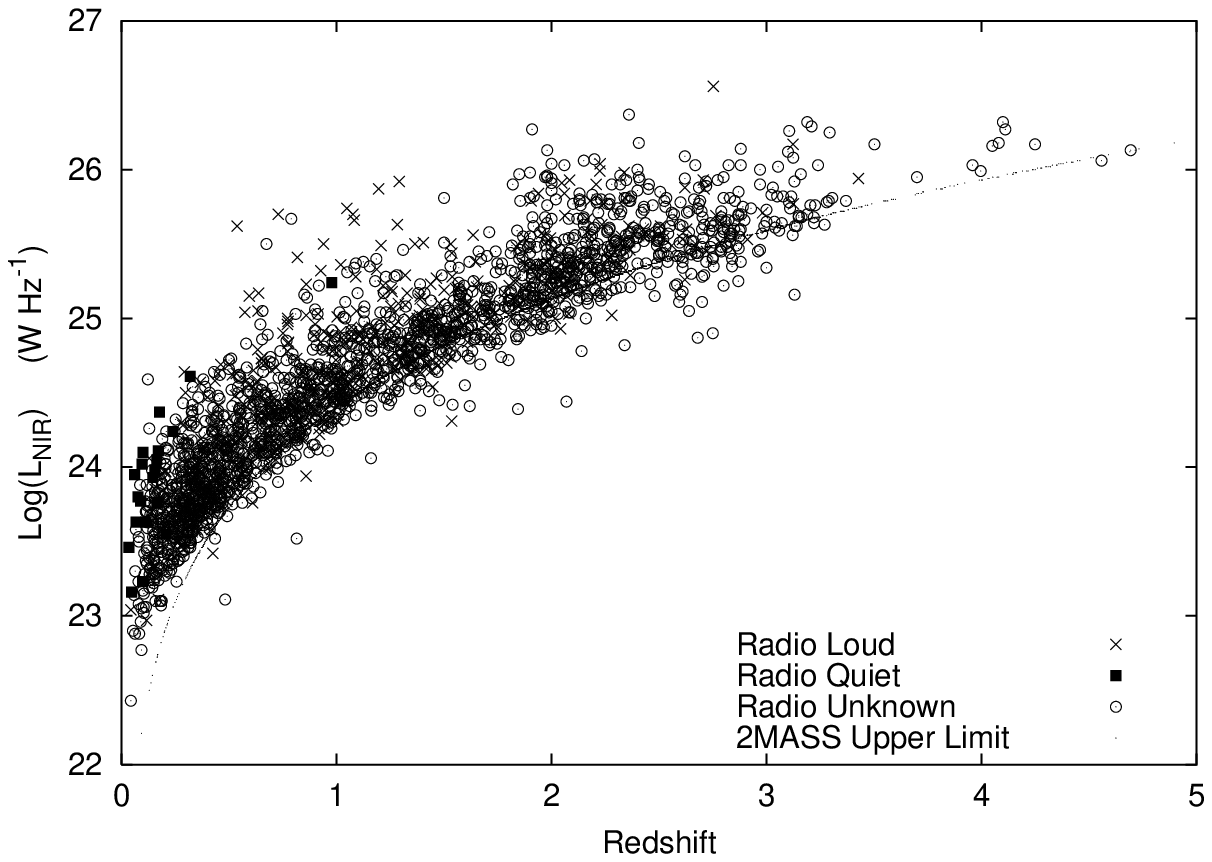}
\caption[]{
This figure shows the correlation between the NIR luminosity and the redshift 
of all quasars in VV00 that fall within the area measured by 2MASS.
RLQs (crosses) and RQQs (filled squares) are plotted as well as quasars which
did not have measured radio flux density in VV00 (``Radio Unknown'';
circled dots).
We have also included quasars which did not have a corresponding detection in 
2MASS (``2MASS upper Limit''; dots) and have used a typical high Galactic
latitude 2MASS
limit of $K= 15.50$ to compute the NIR luminosity.  
}\label{logIRz}
\end{figure}

Figure \ref{logIRz} 
shows the degeneracy between luminosity and redshift inherent in any
single flux-limited survey.  However, it is clear that 2MASS can in principle
detect quasars to at least $z=5$.

\begin{figure}
\plotone{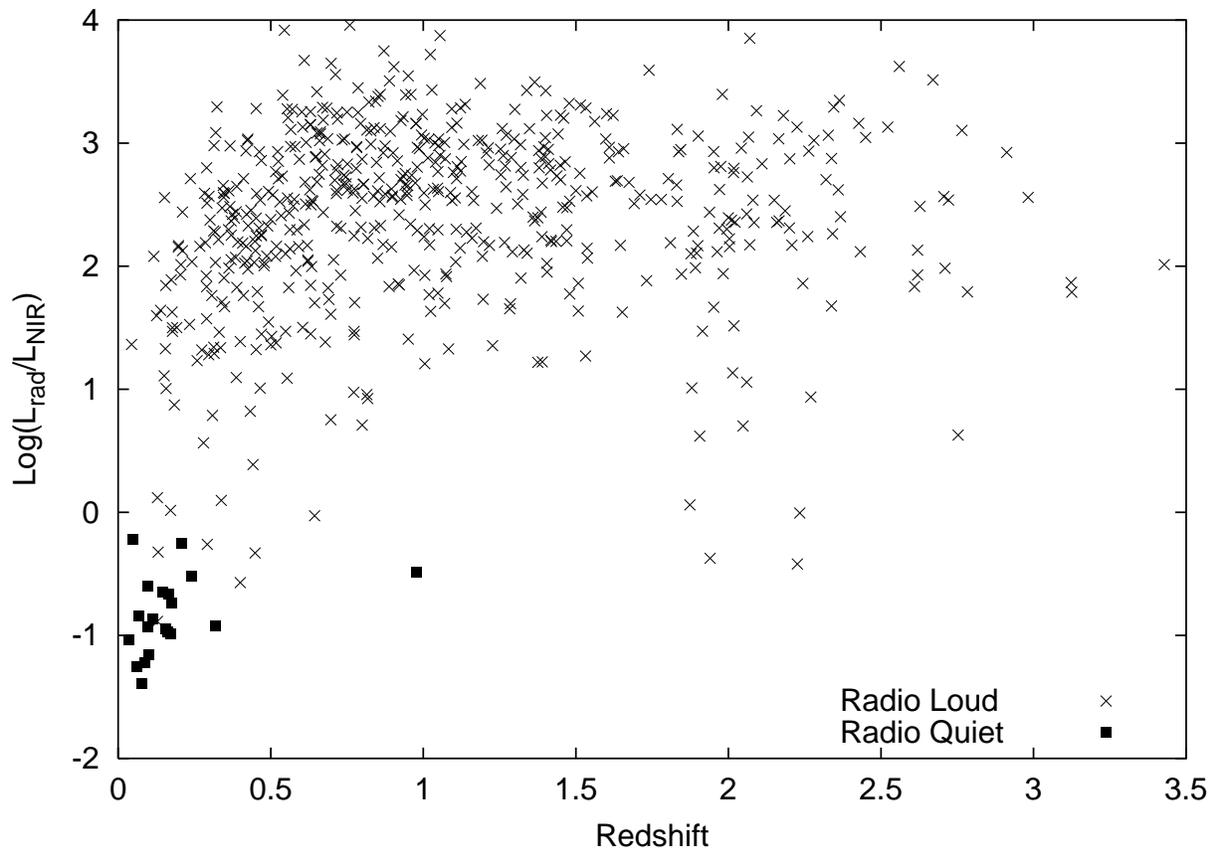}
\caption[]{
This figure shows the relationship between the ratio of $L_{rad}/L_{NIR}$ as a
function of redshift for RLQs (crosses) and RQQs (filled squares).  Most of the
quasars in this plot are radio-loud because most RQQs have only upper limits to
their radio fluxes, which are not reported by VV00.
}\label{logRIRz}
\end{figure}

Figure \ref{logRIRz} 
shows that most VV00 quasars with radio information are radio-loud quasars,
since VV00 do not report upper limits on radio flux densities.
Assuming that VV00 reported radio data for all radio-selected quasars,
the majority of radio-selected quasars catalogued in VV00 are radio-loud,
and most radio-quiet quasars in VV00 must not have been radio-selected.
The RLQ vs. RQQ bimodality seen in heterogeneous quasar samples is not erased
by plotting $L_{rad}/L_{NIR}$ instead of $L_{rad}/L_{opt}$. However, it is
interesting to note that some objects which qualify as radio-loud using
$L_{rad}/L_{opt}$ would be classified as radio-quiet using $L_{rad}/L_{NIR}$ 
(or vice-versa, depending on where the division between RLQ and RQQ is made in
$L_{rad}/L_{NIR}$).  One possible explanation for this might be because some
radio-selected quasars in VV00 are obscured by dust and have 
low values of $L_{opt}$ for their $L_{NIR}$.

\begin{figure}
\plotone{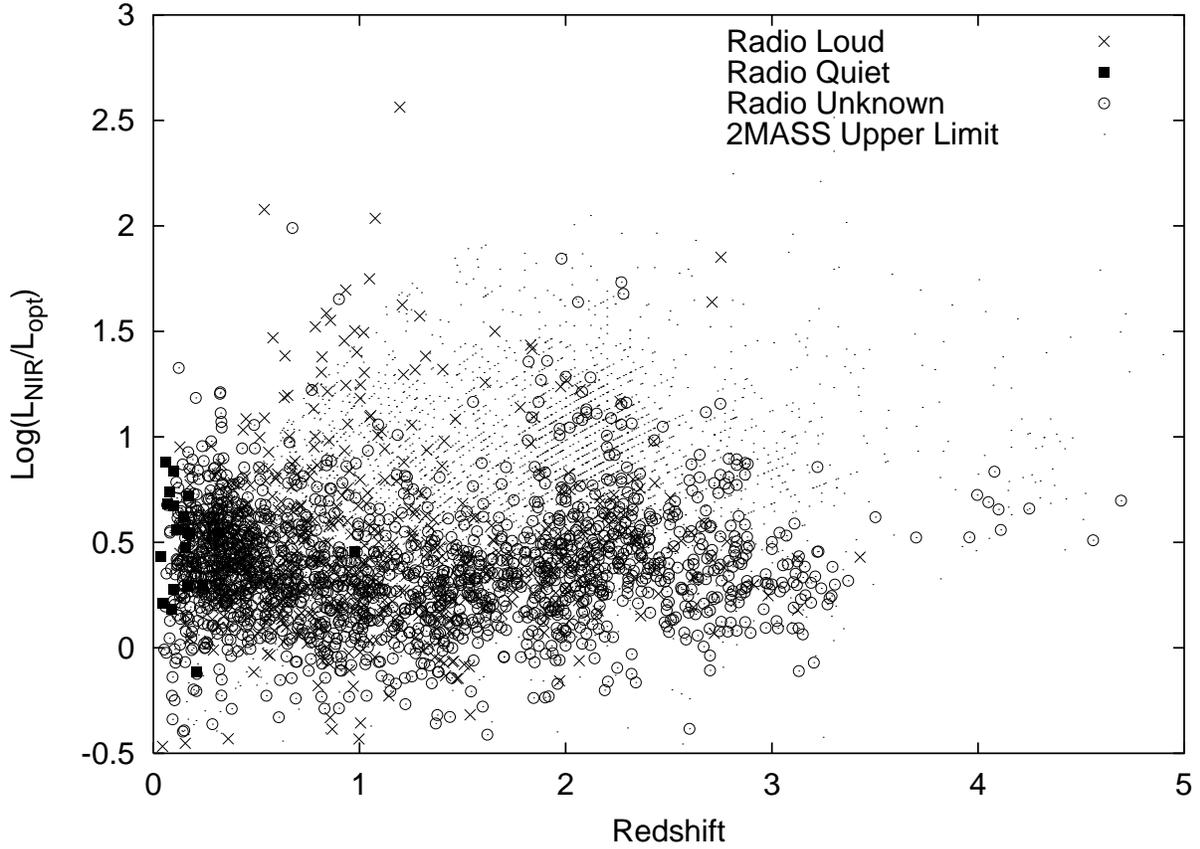}
\caption[]{
This figure illustrates the relationship between the ratio of
$L_{NIR}/L_{opt}$ versus redshift for the sample of quasars in VV00 
which lie within the 2MASS coverage area.  
Symbols are the same as in Figure~\ref{logIRz}.
Radio-loud quasars are more likely to have large $L_{NIR}/L_{opt}$ ratios
than radio-quiet or radio-unknown quasars.
The peak value of the $L_{NIR}/L_{opt}$ distribution is shifted slightly higher
at $z<0.5$ probably because of host galaxy contributions to $L_{NIR}$,
and at $z\sim2.3$ probably because of $H\alpha$ contributions to $L_{NIR}$.
Neither of these effects are compensated for by our simple power-law
$k$-correction.
The handful of $z>3.5$ objects suggest a shift toward larger $L_{NIR}/L_{opt}$
at those redshifts as well, which could be due to a systematic inaccuracy below
the Lyman limit in the $k$-correction applied by VV00 in their
calculation of $M_B$.  Such a shift would not affect our conclusions, however,
given the small number of objects in VV00 at $z>3.5$.
The many upper limits at large $L_{NIR}/L_{opt}$ belong to optically
faint quasars whose $L_{NIR}$ values are poorly constrained by the relatively
shallow 2MASS data.
}\label{logIROz}
\end{figure}

Figure \ref{logIROz} 
shows that radio-loud quasars from VV00 are more likely to have large
$L_{NIR}/L_{opt}$ ratios ($L_{NIR}/L_{opt} \gtrsim 1$) than radio-quiet quasars
from VV00.  9.9\% of known RLQs from VV00 
which are detected in 2MASS have $L_{NIR}/L_{opt} \geq 1$
compared to only 2.5\% of RQQs and quasars with unknown radio properties.
Again, this is understandable if some radio-selected quasars in VV00 
are obscured by dust and consequently have their values of 
$L_{opt}$ suppressed relative to their $L_{NIR}$.
Radio-quiet quasars similarly obscured by dust would be underrepresented in
magnitude-limited optically selected samples, especially color-selected ones,
and thus less likely to be catalogued in VV00.
Note that the many upper limits at large $L_{NIR}/L_{opt}$ belong to optically
faint quasars whose $L_{NIR}$ values are poorly constrained by the relatively
shallow 2MASS data.  Also note the shifts toward slightly larger 
$L_{NIR}/L_{opt}$ at $z<0.5$ and $2.1\lesssim z \lesssim 2.5$.
The former is probably due to host galaxy contributions to $L_{NIR}$,
and the latter to the presence of H$\alpha$ in the $K_s$ band, 
neither of which are compensated for by our simple power-law $k$-correction.

To determine if the $L_{NIR}/L_{opt}$ distribution of the RLQs differed
significantly from that of the other quasars, i.e. the RQQs and quasars with
unknown radio properties (RUQs), we used the Peto-Prentice Generalized Wilcoxon
test for censored datasets \citep{fn85}, as implemented in the 
{\sc stsdas.statistics} package of {\sc iraf}.\footnote{The Image Reduction and 
Analysis Facility (IRAF) is distributed by the National Optical Astronomy
Observatories, which is operated by AURA, Inc., under contract to
the National Science Foundation.}  This test is insensitive to differences in
the censoring patterns for the two samples being compared. 
The combination of RQQs and RUQs
forms a valid, conservative sample for comparison with the RLQs, since RQQs
comprise $\sim$90\% of quasars and since any unrecognized RLQs among the RUQs
will dilute any intrinsic difference between the $L_{NIR}/L_{opt}$ distributions
of RLQs and RQQs.  We find that the $L_{NIR}/L_{opt}$ ratios distributions of
these two samples are different at the 99.99\% confidence level.  Again, we do
not believe that the higher fraction of large $L_{NIR}/L_{opt}$ ratio objects
among {\em known} RLQs compared to {\em known} RQQs is an intrinsic property
of RLQs, but is merely a reflection of the lesser bias against dust-obscured 
quasars with large $L_{NIR}/L_{opt}$ ratios in radio-selected samples
compared to optically-selected samples.

\begin{figure}
\plotone{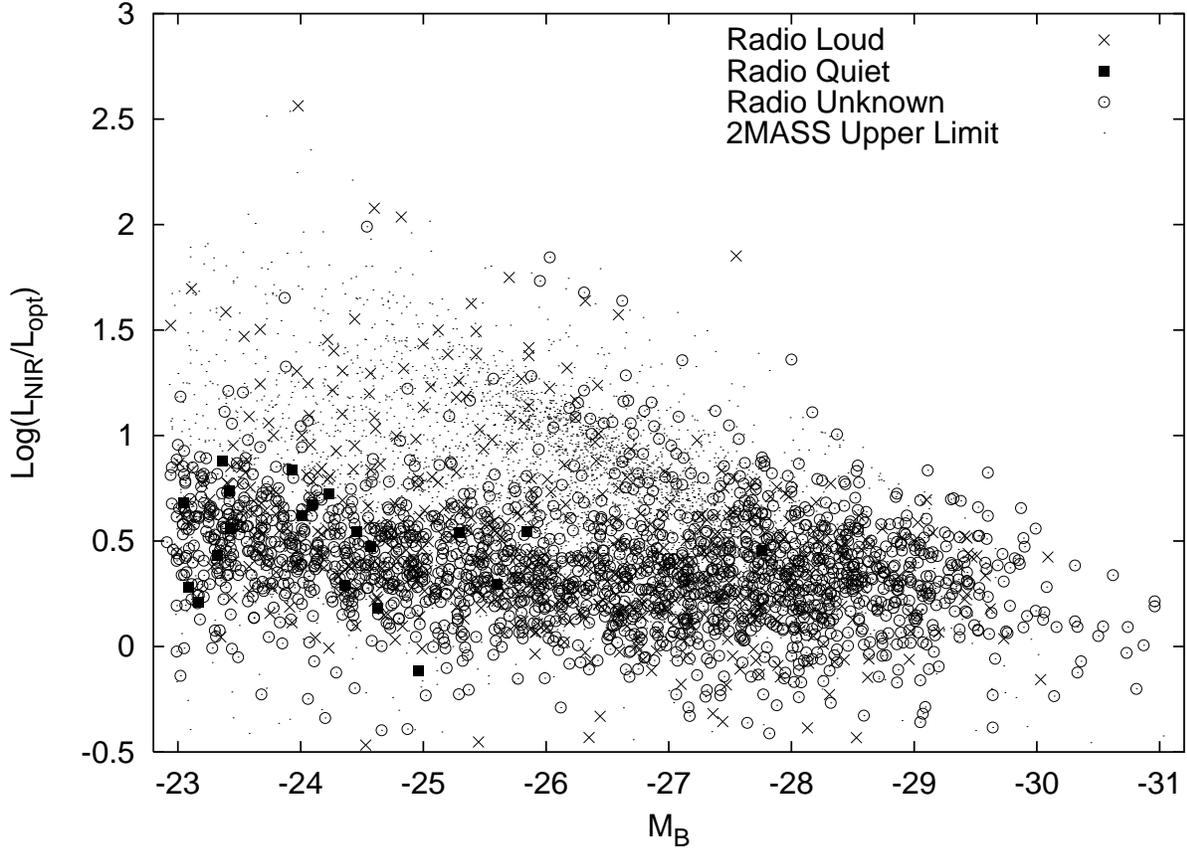}
\caption[]{
The ratio of the near-infrared to optical luminosity is shown as a function of 
absolute B magnitude for the sample of quasars in VV00 which lie within the
2MASS coverage area.  Symbols are the same as in Figure~\ref{logIRz}.
The $M_B$ values from VV00 are accurate only to $\pm0.1$, so a random offset
between $-$0.04 and +0.0.05 magnitudes has been added to them to spread out
the points in the figure.
The peak value of the $L_{NIR}/L_{opt}$ distribution is shifted slightly higher
at low luminosities, probably because of host galaxy contributions to $L_{NIR}$
in the low redshift objects which dominate the sample at low luminosities.
The fact that most quasars with large $L_{NIR}/L_{opt}$ ratios are
fainter than $M_B \simeq -27$ is consistent with the suggestion that 
quasars with large $L_{NIR}/L_{opt}$ ratios are obscured by dust.
The many upper limits at large $L_{NIR}/L_{opt}$ belong to optically
faint quasars whose $L_{NIR}$ values are poorly constrained by the relatively
shallow 2MASS data.
}\label{logIROB}
\end{figure}

Figure \ref{logIROB} 
corroborates the suggestion that quasars with large $L_{NIR}/L_{opt}$ ratios
are obscured by dust.  It shows that such quasars 
are much less common among the population of quasars more luminous
than $M_B \simeq -27$ than among the less luminous population.
The Peto-Prentice test shows that this difference is significant at
$\geq$99.995\% confidence whether the sample of all VV00 quasars is split
at $M_B=-27$ or is split in half (at $M_B=-26.2$).
This is to be expected if dust obscuration is lowering $L_{opt}$ 
(and $M_B$) relative to $L_{NIR}$, thus increasing $L_{NIR}/L_{opt}$.
By itself this does not rule out alternative explanations.
However, host galaxy contributions to the NIR luminosities would require 
extremely luminous host galaxies since the quasars with large
$L_{NIR}/L_{opt}$ ratios have $0.5<z<3$ (Figure~\ref{logIROz}).
Also, a scenario where large $L_{NIR}/L_{opt}$ ratios are caused by synchrotron
emission extending into the NIR in some radio-loud quasars \citep{sr97} 
does not explain why such emission should preferentially occur in 
less luminous quasars.  
%
Dust obscuration seems the simplest explanation for Figure \ref{logIROB},
although we note that Francis, Whiting \& Webster (2000) concluded that 
synchrotron emission was the best explanation for the red colors of the
many RLQs in the Parkes Half-Jansky sample of \citet{web95}.

\section{Selection of Quasars Using 2MASS}\label{Selection}

As discussed in the introduction, 
if there indeed exists a population of dust-obscured radio-loud quasars, then
it is desirable to search for a population of dust-obscured radio-quiet quasars
without being strongly biased against such objects.
Selection on NIR magnitudes reduces bias due to dust extinction,
and selection on NIR colors reduces bias due to dust reddening.
We use the sample of VV00 quasars detected by 2MASS to investigate NIR 
selection.  However, it must be kept in mind that the majority of quasars 
from VV00 are probably not strongly dust-obscured due to their predominantly
optical selection.  The range of quasar colors in our large sample may
reflect the true range of quasar colors, but the distribution of colors 
will not reflect the true distribution.

\subsection{Selection on Near-Infrared Colors}  \label{nircolors}

\begin{figure}
\plotone{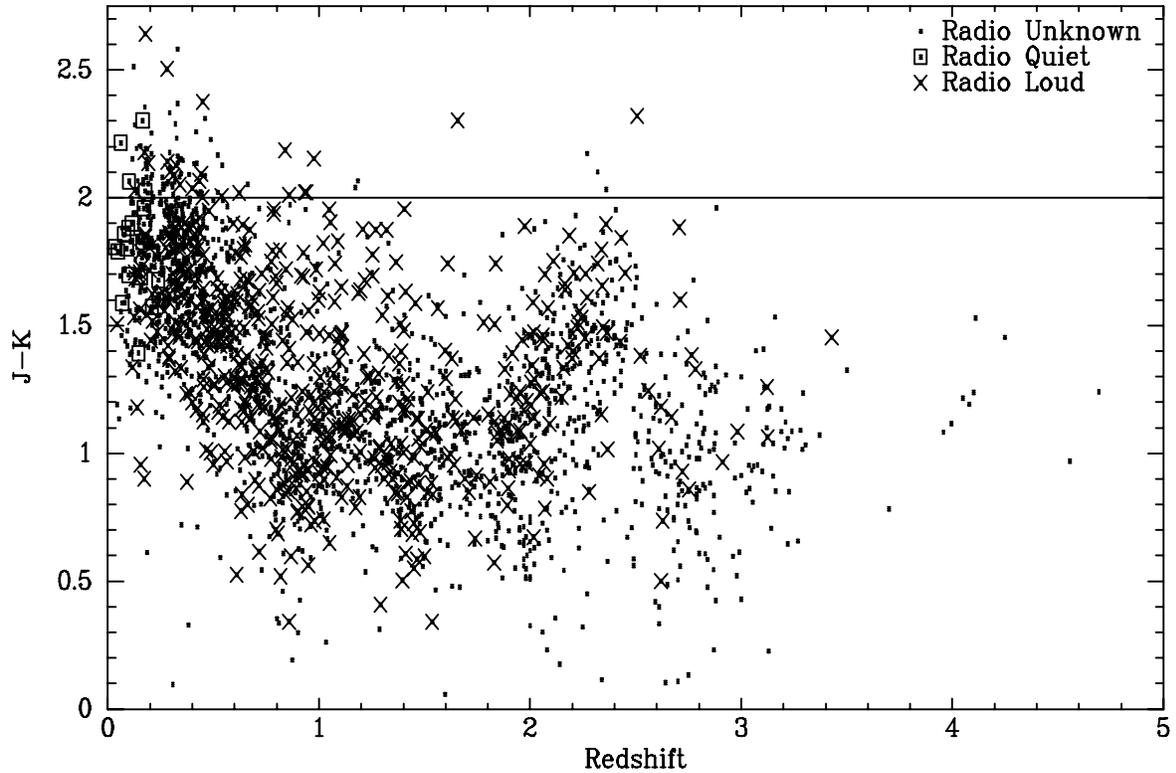}
\caption[]{
$J-K$ versus redshift for all 2MASS counterparts of VV00 quasars 
(small filled squares).
Known radio-loud quasars are denoted by overplotted crosses.
Known radio-quiet quasars are denoted by overplotted boxes; 
they are found only at low redshift because VV00 do not list upper limits
to radio fluxes, and thus more distant radio-quiet quasars form the bulk
of the ``radio unknown" population.
The solid line shows the $J-K>2$ selection criterion of Nelson et al. (1998).
}\label{zjk.epsl}
\end{figure}

\citet{nea98} have begun identifying candidate obscured AGN in 2MASS using a
simple criterion of $J-K\geq2$, as reported in \citet{cut00} and \citet{smi00}.
This criterion was chosen based on the colors of a sample of miscellaneous AGN
including the Palomar-Green (PG) AGN \citep{sg83,nea87},
from which \citet{cut00} conclude that
``virtually all of these known QSOs have $J-K\leq2$...." However, as
seen in Figure \ref{zjk.epsl}, a $J-K\geq2$ criterion selects 13\% of
quasars from VV00 at $z\leq0.5$, as well as a smaller percentage at $0.5<z<3$.
Thus, unless the known quasars with $J-K\geq2$ in VV00 are preferentially 
dust-obscured quasars, a $J-K\geq2$ selection criterion will suffer some 
contamination from unobscured quasars. On the other hand, all but four of the
2277 VV00 quasars detected by 2MASS have $J-K\leq2.5$, whereas the 2MASS AGN of
\citet{cut00} have values as red as $J-K=3.5$. Therefore a selection criterion
of $J-K\geq2.5$ should select dust-obscured AGN with little contamination.

However, a $J-K\geq2.5$ selection criterion would not be complete; that is,
it would not sample the full population of obscured AGN. 
We know this for two reasons.
First, \citet{cut00} do not find evidence of dust reddening only in AGN with
$J-K\geq2.5$.  
Second, we show in the next section that $J-K$ color is not well correlated 
with $B-K$ color, whose longer wavelength baseline than any optical-NIR color 
besides $U-K$ makes it arguably the most efficient photometric measure of the
reddening of a quasar in the optical-NIR \citep{web95}, given the difficult
in obtaining deep wide-field $U$-band data.
A long wavelength baseline is needed to distinguish reddening from the
intrinsic dispersion in quasar colors.
The observed dispersion in the 2MASS quasar sample is only 0\fm31 in $J-K$,
vs. 0\fm67 in $B-K$ (calculated using $5\sigma$ clipping to remove
outliers, and using only $1<z<2$ quasars for $J-K$).
However, the sensitivity to reddening is given
by the ratio of the color excess to this dispersion:
$E(J-K)$/$\sigma_{J-K}$=2.12 vs. $E(B-K)$/$\sigma_{B-K}$=3.84 at $z=3$
(the difference is even more extreme at $z=0$).
Simply put, less extinction is required to produce anomalously red colors in
$B-K$ than in $J-K$, 
or indeed than in any other optical-NIR color besides $U-K$.
Finally, the variation of
$J-K$ color with redshift seen in Figure \ref{zjk.epsl} means that a
simple $J-K$ color cut will not select reddened AGN with equal efficiency
at all redshifts. Thus the evidence available to date, while scarce,
suggests that infrared colors alone are insufficient to {\em efficiently} 
sample the full parameter space open to obscured quasars.

\subsection{Selection on Optical and Near-Infrared Colors}\label{optcolors}

\begin{figure}
\plotone{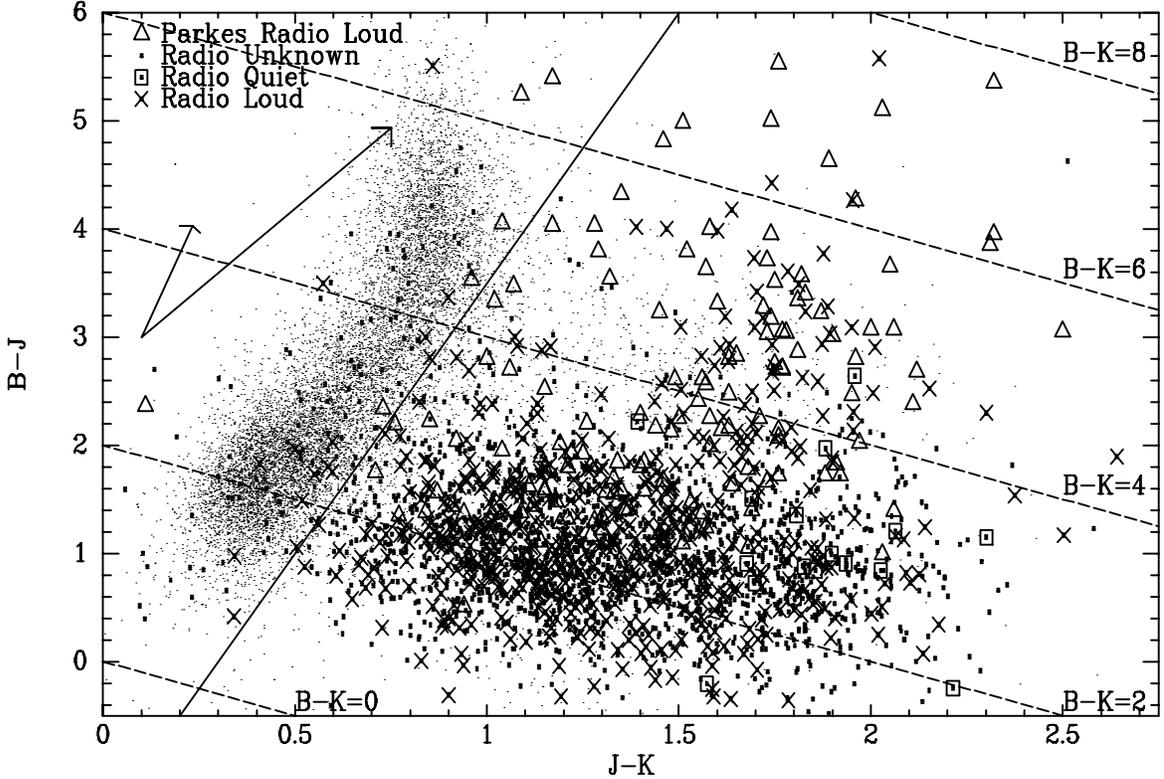}
\caption[]{
$B-J$ vs. $J-K$ color-color diagram for all 2MASS counterparts of VV00 quasars.
The small dots are approximately 14,000 essentially random $K\leq15$ 
2MASS point sources.
They illustrate the colors of the stellar locus as well as other objects 
(quasars and compact galaxies) seen at moderate to high Galactic latitude.
The small filled squares are the 2277 VV00 quasars with 2MASS counterparts 
within 4\arcsec.
Known radio-quiet quasars are denoted by overplotted boxes, 
and known radio-loud quasars are denoted by overplotted crosses.
Dashed lines show lines of constant $B-K$ color.
Radio-loud quasars outnumber radio-quiet or radio-unknown quasars for
$B-K\gtrsim3.5$, whereas the opposite is true for $B-K\lesssim3.5$.
This heterogeneous sample of RLQs from VV00 has a range of colors similar to the
well-defined Parkes sample of Francis, Whiting \& Webster 
(2000; open triangles).
The two arrows show the reddening vectors for one magnitude of visual
extinction in the quasar rest frame
($A_V=1$, or $E(B-V)=0.32$; Table 21.6 of Allen 2000).
The shorter arrow is for $z=0$ and the longer arrow is for $z=3$.
The solid line shows the selection criteria we adopt to separate stars from
quasar candidates; note that reddening moves quasars approximately parallel
to the selection criteria.
}\label{jkbjplusparkes.epsl}
\end{figure}

The 2MASS catalog includes $B$ and $R$ magnitudes from the USNO-A and 
Tycho catalogs, as well as $JHK$. 
As noted by \citet{fww00} and \citet{cut00}, sampling the full parameter 
space open to obscured quasars will be difficult based on $B-R$ vs. $J-K$
colors.  However, \citet[hereafter WHF]{whf00} have proposed a promising
criterion for selecting $K$-band flux-limited samples of quasars from the
$V-J$ vs. $J-K$ color-color diagram, which they dub the KX method.
We do not have V magnitudes for the majority of our quasars, but the
$B-J$ vs. $J-K$ diagram (Figure \ref{jkbjplusparkes.epsl}) shows the same
gross features as the $V-J$ vs. $J-K$ diagram of WHF (their Figure 2).
Note that we plot our diagram upside-down relative to theirs, however.
The dotted points in Figure \ref{jkbjplusparkes.epsl} are 
an essentially random sample of 14,000 high Galactic latitude objects.
The filled squares are the 2277 VV00 quasars with 2MASS counterparts within 
4\arcsec:  known radio-quiet quasars among them are denoted by overplotted 
boxes, and known radio-loud quasars by overplotted crosses.
The bulk of known quasars have blue $B-J$ colors, but there is a tail of 
candidate obscured quasars, predominantly RLQs, with $J-K\sim1.75$ and
$B-J\geq2$.  Note that most of these quasars do not have $J-K>2$. 
The dashed lines of constant $B-K$ color show that red $J-K$ color 
does not correlate strongly with red $B-K$ color either.

We cannot rule out the possibility that $J-K$ does correlate 
with $B-J$ or $B-K$ but that this correlation is masked by
strong selection effects against quasars with red
$B-J$ and $J-K$ in Figure \ref{jkbjplusparkes.epsl}. However, we have
also plotted a subsample of radio-selected quasars from the completely
identified Parkes Half-Jansky sample \citep[triangles;][]{fww00}.
There is no strong
correlation between $J-K$ and $B-J$ or $B-K$ for these objects either.
Selecting objects red in $J-K$ would not appear to be an efficient way
of selecting objects red in $B-K$.
With RLQs, there is no guarantee that such red $B-K$ colors are due to dust 
rather than synchrotron emission with a short wavelength cutoff in the
rest-frame optical \citep{sr97,fww00}. However, with RQQs the latter effect 
will be much weaker and any confirmed population of RQQs with $B-K\gtrsim4$ 
will almost certainly be reddened by dust.

\begin{figure}
\plotone{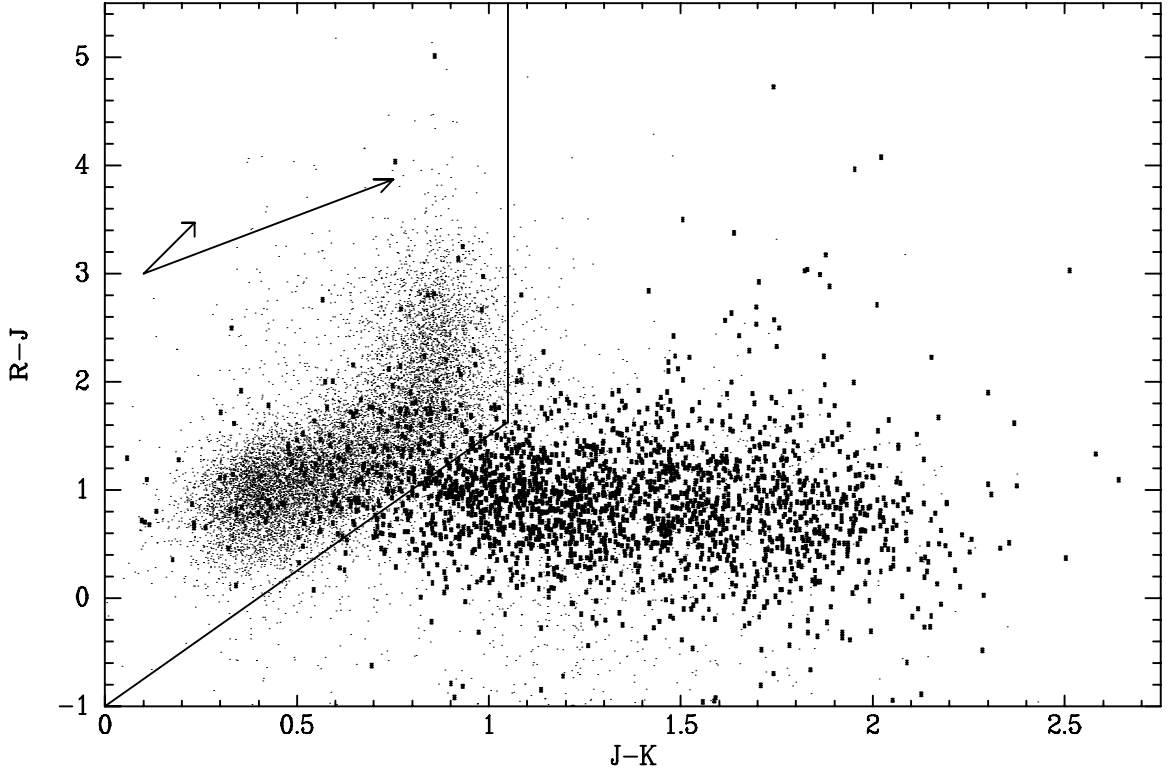}
\caption[]{
$R-J$ vs. $J-K$ color-color diagram for all 2MASS counterparts of VV00 quasars.
The small dots are approximately 14,000 essentially random $K\leq15$
2MASS point sources.
The small filled squares are the 2277 VV00 quasars with 2MASS counterparts
within 4\arcsec.  
The open triangles are the Parkes quasars of Francis, Whiting \& Webster (2000).
The two arrows show the reddening vectors for one magnitude of visual
extinction in the quasar rest frame
($A_V=1$, or $E(B-V)=0.32$; Table 21.6 of Allen 2000).
The shorter arrow is for $z=0$ and the longer arrow is for $z=3$.
The solid line shows the selection criteria we adopt to separate stars from
quasar candidates; note that reddening moves quasars approximately parallel
to the selection criteria.
}\label{jkrjplusparkes.epsl}
\end{figure}

The $R-J$ vs. $J-K$ color-color diagram (Figure \ref{jkrjplusparkes.epsl}) 
is similar to the $V-J$ and $B-J$ vs. $J-K$ diagrams, and as pointed out
by WHF offers the advantage of extending the effectiveness of the KX method
from $z\simeq3.4$ to $z\simeq4.4$. Beyond $z\simeq4.4$, Ly$\alpha$ forest
absorption greatly depresses the flux in the $R$ band, and the resulting
red $R-J$ colors may result in confusion with stars.

We have constructed selection criteria similar to the $V-J/J-K$ criterion of
WHF to illustrate the separation of quasars from stars in both $B-J/J-K$ and
$R-J/J-K$ color-color diagrams (solid lines in Figure \ref{jkbjplusparkes.epsl}
and Figure \ref{jkrjplusparkes.epsl}).
We adopt the following selection criteria in $B-J/J-K$:
\begin{eqnarray}
(J-K) \geq 0.3 + 0.2 \times (B-J)
\end{eqnarray}
and the following in $R-J/J-K$:
\begin{eqnarray}
(J-K) \geq 0.4 + 0.4 \times (R-J)\ \ ({\rm for\ } R-J < 1.625) \\
(J-K) \geq 1.05\ \ ({\rm for\ } R-J \geq 1.625) 
\end{eqnarray}
As pointed out by WHF and as shown by the arrows in 
Figure \ref{jkbjplusparkes.epsl} and Figure \ref{jkrjplusparkes.epsl},
reddening moves quasars approximately parallel to 
these selection criteria and the bluer portion of the stellar locus in these
diagrams, resulting in good sensitivity to reddened and unreddened quasars.
Only 124 of 2277 quasar counterparts (5.4\%) have colors that do not meet
the $B-J/J-K$ selection criteria, while 283 (12.4\%) have colors that do
not meet the $R-J/J-K$ selection criteria. Some of these are probably
chance coincidences of stars with VV00 quasars. Assuming all 37 estimated
spurious matches are among them, then only 87 quasar counterparts (3.8\%)
do not meet the $B-J/J-K$ selection criteria while 236 quasar counterparts
(10.4\%) do not meet the $R-J/J-K$ selection criteria.
(Since these criteria are illustrative only, we do not consider how
efficient they are at rejecting stars, but any application of them to
2MASS or other data must consider how close to the stellar locus the
selection criteria can be drawn as a function of magnitude without 
selecting too many spurious candidates.)

\begin{figure}
\plottwo{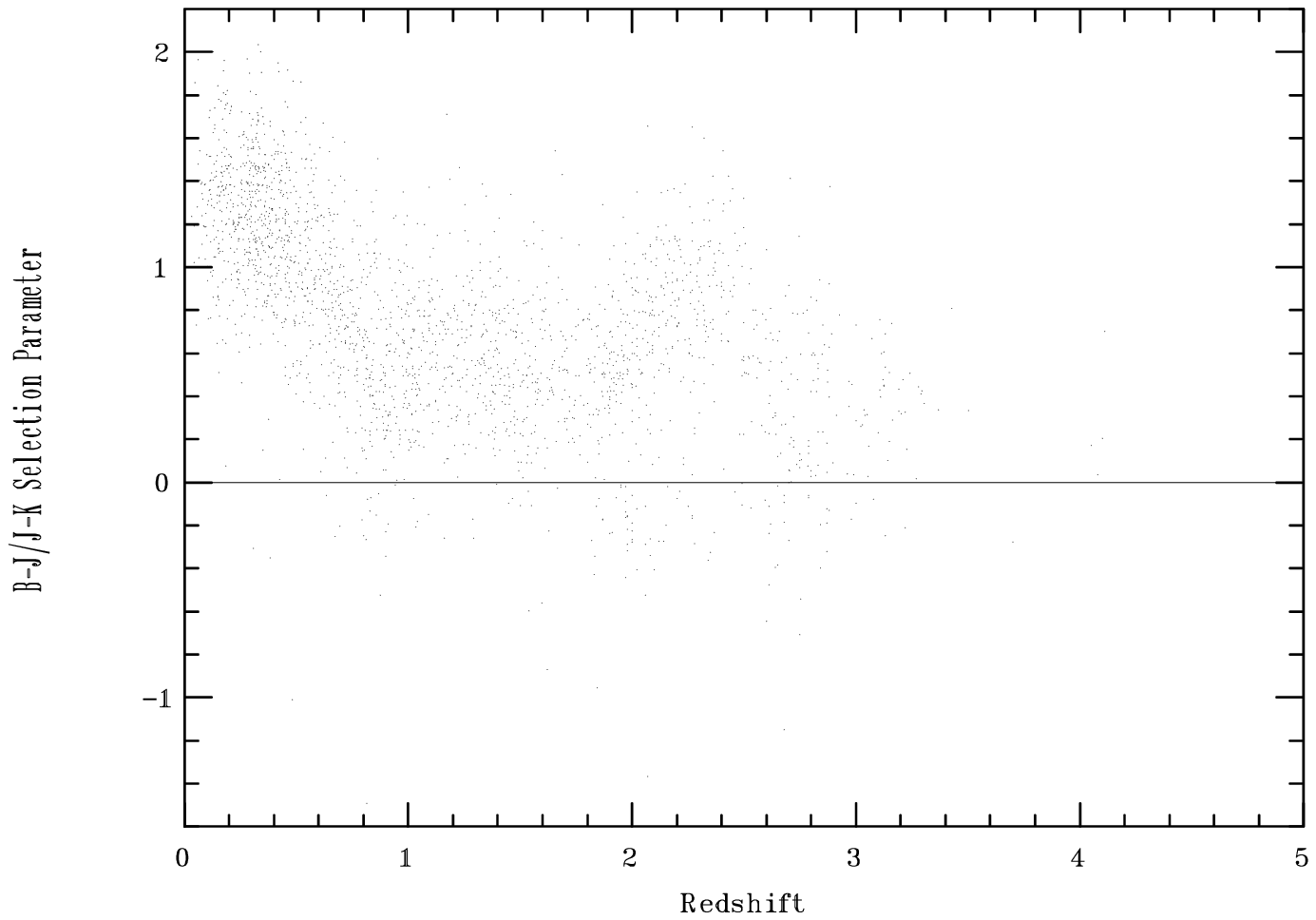}{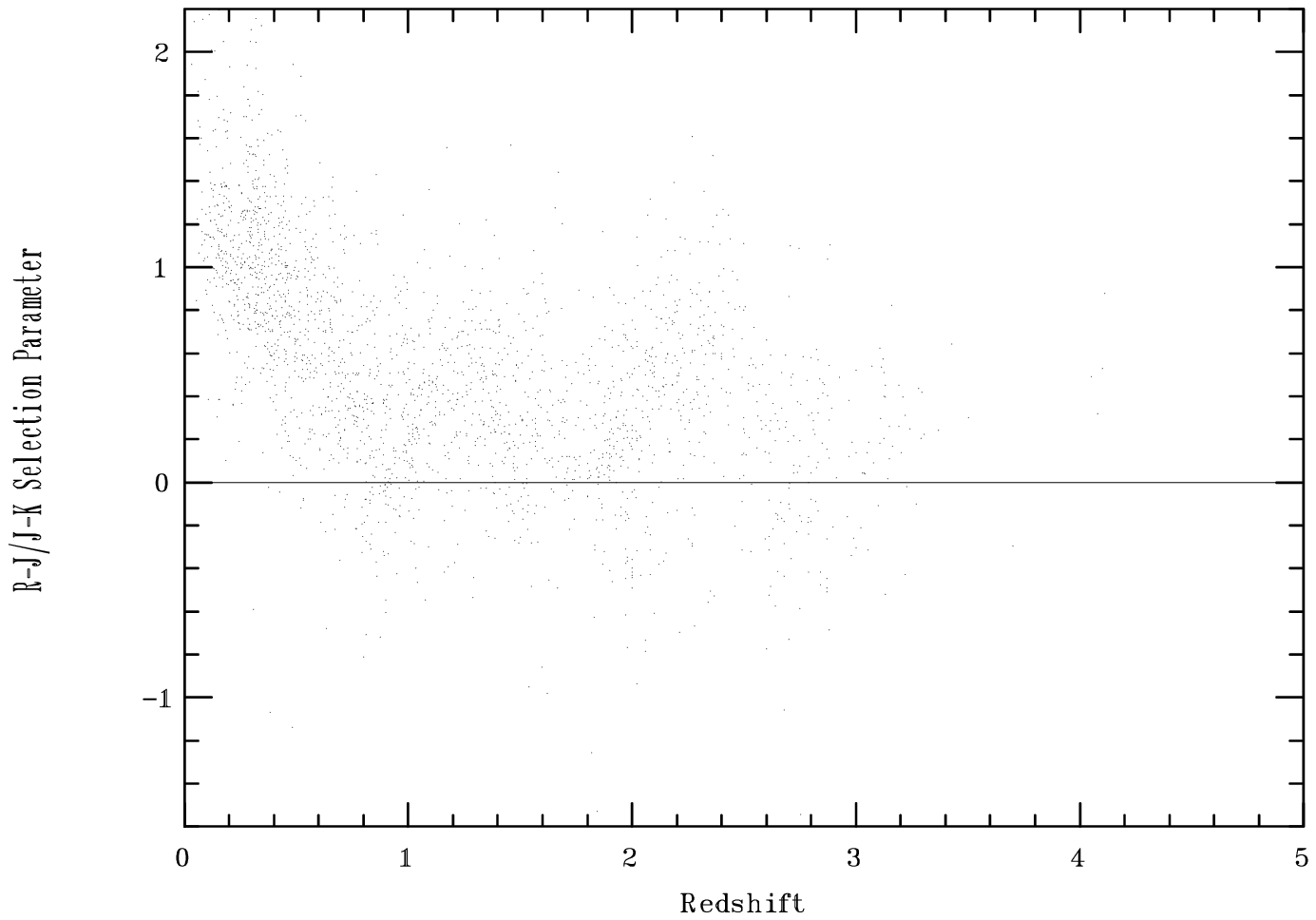}
\caption[]{
Quasar selection criteria parameters  
versus redshift for all 2MASS counterparts of VV00 quasars.  
In a) we show the $B-J/J-K$ selection parameter $(J-K)-0.3-0.2\times(B-J)$,
and in b) we show the analogous parameter for $R-J/J-K$ selection.
In both cases this parameter is the shortest distance in color-color space
to the selection criterion line which divides quasar candidates from stars.
Objects with positive values of these quantities (above the solid lines)
would be accepted as quasar candidates.
The intrinsic dispersion and redshift dependence of quasar colors means that
the selection efficiency will be redshift dependent,
but overall $B-J/J-K$ selection is more efficient than $R-J/J-K$ selection.
}\label{zKX.epsl}
\end{figure}

For both selection criteria, the selection efficiency is dependent on 
redshift.  Figure \ref{zKX.epsl}a shows our $B-J/J-K$ selection parameter
versus $z$ for all VV00 quasar counterparts, while
Figure \ref{zKX.epsl}b shows our $R-J/J-K$ selection parameter.
The selection parameter is essentially the distance in the color-color diagram
to the selection criterion line which divides quasar candidates from stars.
Objects with a positive selection parameter are accepted as candidates
from that diagram.  
While $R-J/J-K$ selection may be more useful at $z\geq3.4$ than $B-J/J-K$
selection due to Ly$\alpha$ forest absorption in $B$, it is less efficient 
at separating quasars from stars at $z\leq3.4$.  However, our $B-J/J-K$ 
criterion is not perfect: the intrinsic dispersion in quasar colors
plus the redshift dependence of quasar colors may result in reduced
selection efficiency at $1<z<2$ compared to $z<0.5$ and $2<z<2.5$.

Thus, the $V-J/J-K$ criterion of WHF, or a $B-J/J-K$ selection criterion 
patterned after theirs, appear to be a more complete method of selecting 
reddened AGN than the simple $J-K>2$ selection of \citet{nea98}. Nevertheless,
\citet{pss00} find that the 2MASS AGN of \citet{nea98} span a much larger
range in $B-K$ color and in optical polarization than the Palomar-Green
(PG) AGN 
even when the comparison is restricted to
PG AGN with $J-K>2$. Both of these trends are consistent with a much higher
frequency of dust-obscured AGN in the 2MASS $J-K>2$ selected sample.  This high
frequency of dust-obscured AGN found using a simple and probably incomplete
criterion is unexpected given the lack of a correlation between $J-K$ and $B-J$
(or $B-K$), but it does suggest that more complete $B-J/J-K$ selected AGN
samples will be very efficient at finding great numbers of dust-obscured AGN.
Of course, a key test will be to see where the 2MASS AGN of \citet{nea98}
fall in the $B-J/J-K$ diagram.

\section{Conclusions}\label{Conclusions}

We have identified near infrared counterparts within $4\arcsec$
of 2277 quasars from the 
Veron-Cetty \& Veron (2000) catalog using the 2MASS Second Incremental 
Data Release. Approximately 6320 
quasars from Veron-Cetty \& Veron (2000) 
were located within the area covered by the 2MASS catalog. 
This detection rate of 36\% suggests that the final 2MASS catalog
will yield matches for approximately 4760 of the 13214 VV00 quasars.
Our sample of 2277 quasars with IR data is nearly ten times larger than
the largest previous such sample \citep{sk97}, and more than ten times larger 
than the largest previous samples with homogeneous IR data \citep{whf00,fww00}.

The observation that the fraction of radio-loud quasars with large
$L_{NIR}/L_{opt}$ ratios is higher than the fraction of probable radio-quiet
quasars with such ratios supports the suggestion that {\em known} 
radio-selected quasars are more likely to be obscured by dust than {\em known} 
optically-selected quasars due to stronger selection effects
against dust-obscured quasars in optically selected samples.
Known quasars with large $L_{NIR}/L_{opt}$ ratios are also underrepresented
among optically luminous quasars ($M_B < -27$).  This is again consistent with
a selection effect caused by dust obscuration, which would tend to lower
$L_{opt}$ and $M_{B}$ while increasing $L_{NIR}/L_{opt}$. 
These results need to be confirmed using homogeneous quasar samples rather
than the heterogeneous VV00 sample, but the statistical significance of our
results in such a heterogeneous sample bodes well for their reality in more
carefully selected samples.

Following Warren, Hewett \& Foltz (2000),
we show that the $B-J/J-K$ color-color diagram is more 
sensitive to the detection of dust-obscured quasars than using infrared
colors alone.  The stage is now set for the determination of the true fraction
of dust-obscured RQQs using large NIR surveys such as 2MASS and DENIS.
The detection of dust-obscured AGN will also benefit greatly from hard X-ray
surveys, but for the foreseeable future NIR selection will be possible over
much larger areas than hard X-ray selection, and thus more sensitive to 
luminous AGN which are more easily studied in detail.




\acknowledgments
We thank {\v Z}eljko Ivezi{\' c} and the referee Paul Francis for useful 
comments, and Roc Cutri for determining which quasars were observed but 
not detected by 2MASS.
This publication makes use of data products from the Two Micron All Sky Survey,
which is a joint project of the University of Massachusetts and the Infrared
Processing and Analysis Center/California Institute of Technology, funded by
the National Aeronautics and Space Administration and the National Science 
Foundation.


\clearpage



\clearpage

\begin{deluxetable}{llllcclllllllllcll} 
\tabletypesize{\tiny}
\tablecaption{VV00 Quasars Detected in 2MASS\tablenotemark{a}
\label{tab_detected}}
\rotate
\tablenum{1}
\tablecolumns{18}
\tablewidth{0pt}
\tablehead{
\colhead{2MASS ID}&\colhead{VV00 ID}&\colhead{RA}&\colhead{DEC}&\colhead{$L_{NIR}$}&\colhead{$L_{opt}$}&\colhead{b\_m}&\colhead{r\_m}&\colhead{J}&
\colhead{$\sigma_{J}$}&\colhead{H}&\colhead{$\sigma_{H}$}&\colhead{K}&
\colhead{$\sigma_{K}$}&\colhead{$M_{B}$}&\colhead{dist\_opt}&\colhead{Radio}
&\colhead{z}\\
\colhead{}&\colhead{}&\colhead{deg}&\colhead{deg}&\colhead{Jy}&\colhead{Jy}&\colhead{mag}&\colhead{mag}&\colhead{mag}&\colhead{mag}&\colhead{mag}
&\colhead{mag}&\colhead{mag}&\colhead{mag}&\colhead{mag}&\colhead{arcsec}&\colhead{Jy}&\colhead{}}

\startdata
0000029-350332&MS~23574-3520&0.012084&-35.059052&24.14&23.91&17.20&17.10&
16.221&0.110&15.332&0.119&14.513&0.099&-25.50&0.14&-99.00&0.508\\
0000244-124548&PHL~2525&0.101767&-12.763336&24.06&23.71&15.10&14.80&14.475&
0.034&13.704&0.041& 12.580&0.034&-25.00&0.31&-99.00&0.200\\
0001323+211336&TEX~2358+209&0.384921&21.226782&24.89&23.87&18.20&18.60&
16.177&0.102&15.547&0.142&14.526&0.093&-25.40&0.25&26.93&1.106\\
0002224-000444&PB~5694&0.593673&-0.078939&24.08&23.39&18.60&18.30&16.586&
0.122&16.500&0.259&15.775&0.229&-24.20&0.79&-99.00&0.810\\
0003153-275157&HE~0000-2808&0.813904&-27.865856&24.68&24.55&17.60&16.50&
15.996&0.091&15.653&0.160&14.944&0.131&-27.10&0.43&-99.00&1.051\\

\enddata
\tablenotetext{a}{The complete version of this table is in the electronic
edition of the Journal or at 
www.astro.utoronto.ca/$\sim$barkhous/quasar\_table1.gz. 
The printed edition contains only a sample.}
\tablecomments{VV00 IDs are the Name entries of VV00.
A handful of objects which have blank Name entries in VV00
are identified here by the prefix VV followed by their coordinates in VV00.\\
Objects having two 2MASS counterparts within 4\arcsec are identified by an 
asterisk preceding their 2MASS ID number. Only the closest counterpart is
listed here.\\
Columns 3 and 4 give the coordinates for J2000. Columns 5 and 6 lists the 
logarithm of the calculated NIR and optical luminosity. Column 16 gives
the distance to 
the associated optical source. Column 17 lists the radio flux density 
measured at 6 cm.\\
Data for columns 2, 15, 17, and 18 are from VV00, while the remaining data
are from 2MASS.\\
Unavailable data entries are given as -99.00.}
\end{deluxetable}

\clearpage

\begin{deluxetable}{llllll}
\tablecaption{VV00 Quasars With Only Upper Limits in 2MASS\tablenotemark{a}
\label{tab_missing}}
\tabletypesize{\scriptsize}
\tablewidth{0pt}
\tablenum{2}
\tablecolumns{6}
\tablehead{
\colhead{VV00 ID}&\colhead{RA}&\colhead{DEC}&\colhead{$L_{opt}$}&\colhead{Radio}
&\colhead{z}\\
\colhead{}&\colhead{deg}&\colhead{deg}&\colhead{Jy}&\colhead{Jy}&\colhead{}}
\startdata
PKS~2357-326&0.084167&-32.350278&24.03&27.53&1.275\\
PKS~2358-161&0.272500&-15.852222&24.67&28.01&2.044\\
TEX~2358+189&0.285833&19.242778&24.63&28.30&3.100\\
UM~195&0.442500&-1.992778&23.71&-99.00&0.867\\
UM~196&0.458333&-1.994444&24.91&-99.00&2.810\\
\enddata
\tablenotetext{a}{The complete version of this table is in the electronic
edition of the Journal or can be found at 
www.astro.utoronto.ca/$\sim$barkhous/quasar\_table2.gz.
The printed edition contains only a sample.}
\tablecomments{IDs are the Name entries of VV00. A handful of objects
which have blank Name entries in VV00 are identified here by their
coordinates and the prefix VV.\\
Columns 2 and 3 give the coordinates for J2000. Columns 4 lists the 
logarithm of the calculated optical luminosity. Column 5 gives the radio 
flux density measured at 6 cm.\\
Unavailable data entries are given as -99.00.}
\end{deluxetable}

\end{document}